\documentclass[aps,twocolumn,superscriptaddress,showpacs,amsfonts,amsmath,amsthm,amssymb,mathtools,latexsym,amssymb,amsthm,nofootinbib]{revtex4-1}
\usepackage[colorlinks,linkcolor=blue,urlcolor=blue,citecolor=blue,bookmarks,bookmarksnumbered]{hyperref}
\usepackage{breqn}
\usepackage{bm,braket,dcolumn}
\usepackage{slashed}
\usepackage{graphicx,subfigure}
\usepackage{epstopdf,epsfig}
\usepackage{enumitem}
\usepackage{braket}
\usepackage{bbm}
\usepackage{autobreak}
\usepackage{verbatim}
\usepackage{float}

\usepackage[usenames,dvipsnames]{color}         
\definecolor{purple}{rgb}{0.5,0,0.5}

\newcommand{\zz}{\mathbb{Z}}
\newcommand\defeq{\mathrel{\stackrel{\makebox[0pt]{\mbox{\normalfont\tiny def}}}{=}}}
\newcommand{\Tr}{\operatorname{Tr}}
\newcommand{\refeq}[1]{Eq.~\eqref{#1}}

\DeclareMathOperator{\Ima}{Im}
\newcommand{\sumtop}{\sum\limits_{\text{\scriptsize top}}}
\newcommand{\TEE}{S_{\text{\scriptsize   top}}}
\newcommand{\la}{\lambda_A}
\newcommand{\lb}{\lambda_B}

\graphicspath{{pics/}{}}

\begin{document}
\newcommand*{\PITT}{Department of Physics and Astronomy, University of Pittsburgh, Pittsburgh, Pennsylvania 15260, United States}\affiliation{\PITT}
\newcommand*{\PQI}{Pittsburgh Quantum Institute, Pittsburgh, Pennsylvania 15260, United States}

\title{Finite-temperature topological entanglement entropy for CSS codes}
\author{Zhi Li}\affiliation{\PITT}\affiliation{\PQI}
\author{Roger S. K. Mong}\affiliation{\PITT}\affiliation{\PQI}

\begin{abstract}
We consider topological entanglement entropy (TEE) at finite temperature for CSS codes, which include some ordinary topological-ordered systems such as the toric code and some fracton models such as the Haah's code and the X-cube model. We find, under the assumption that there is no extended critical phase, the finite-temperature TEE is a piecewise constant function of the temperature, with possible discontinuities only at phase transitions. We then consider phase transitions of CSS codes.  We claim that there must exist a phase transition at zero temperature for any CSS codes in 2D and 3D with topological order. This statement can be rigorous proved for some familiar examples, while for general models it can be argued based on the low-temperature expansion. This indicates the break down of topological orders at finite temperature. We also discuss possible connections with self-correcting quantum memory.
\end{abstract}
\maketitle

\section{Introduction}
Topological-ordered many body systems, characterized by topological ground state degeneracy, anyon excitations, braiding and fusions, are one of the most important topic in condensed matter physics. It has been proposed that topological-ordered systems can be utilized to realize fault tolerant quantum computation \cite{Kitaev-TQC}.

Topological orders are intimately related to long range quantum entanglement \cite{Chen-Gu-Wen}. For two-dimensional topological-ordered systems, it is found that \cite{Kitaev-tee,Wen-tee} the entanglement entropy contains a constant term, called topological entanglement entropy (TEE), which is related to other characterization of topological orders like the quantum dimension.

Although topological orders are stable against local perturbations and disorders \cite{Wen-Niu}, it might not be stable against thermal fluctuations. This problem is important since any topological quantum computer in real life is subjected to a finite temperature. For example, it is rigorously proved that the 2D toric code is thermally unstable \cite{Alicki-2D}. Moreover, a No-Go theorem \cite{Nogo-2D} claims that string-like logical operators are unavoidable in 2D, indicating the thermal instability. On the other hand, the 4D toric code is thermally stable \cite{Alicki-4D}. The question in three dimension, which is physically more relevant, is still not finally concluded. A No-Go theorem in 3D \cite{Yoshida-nogo,Haahmath} confirms the existence of string-like logical operators for translational invariant systems if the ground state degeneracy is independent of system size. However, one can bypass the condition of this theorem with fractonic systems \cite{Haahcode}. See \cite{review1,review2} for reviews regarding thermal (in)stability for quantum memories.

Fracton topological orders \cite{fracton1,fracton2,Haahcode,Yoshida-fractal,fracton3,FSduality,couplelayer,gaugefracton,rahul} are new kinds of topological phases in 3D characterized by immobile or subdimensional excitations and ground state degeneracy that grows with system size. Due to the restricted mobility of the excitations and the absence of string-like logical operators, one may expect that fractonic systems behave better \cite{Haah-relaxation} against thermal fluctuations than ordinary topological-ordered systems.

\begin{figure}[htp]
	\centering
	\includegraphics[width=\columnwidth]{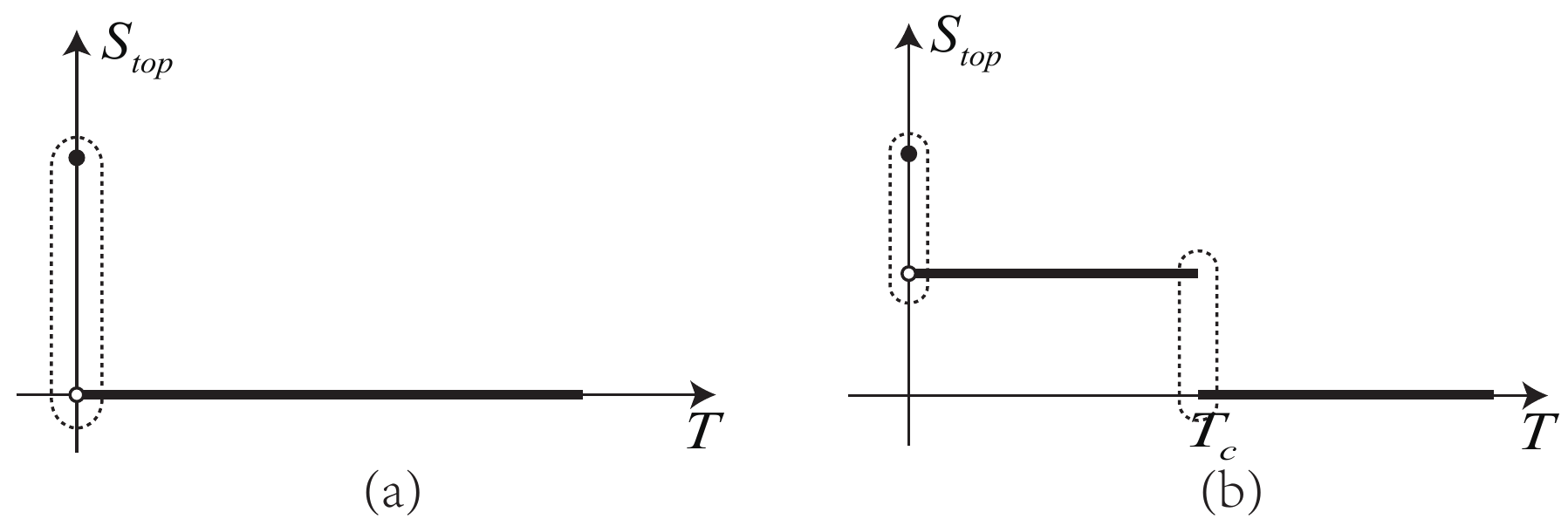}
	\caption{$\TEE$ as a function of temperature. Dotted circle indicates a drop. (a) For 2D toric code, $\mathrm{TEE}=0$ for any finite temperature. (b) For 3D toric code, TEE drops a half at $T=0^+$, remains constant when $0<T<T_c$, then drops to zero at $T=T_c$. We will see Haah's code and X-cube model have similar behavior as (a).}		\label{pic-piecewise}
\end{figure}

From the quantum entanglement point of view, one can consider the topological entanglement entropy at finite temperature~\cite{Chamon2d,Chamon3d,Hamma}. For the 2D toric code, the TEE was computed exactly and it was shown that $\mathrm{TEE}=0$ for any finite temperature $T$, see Fig.~\ref{pic-piecewise}(a). This agree with the thermal instability for topological orders in 2D. On the other hand, for the 3D toric code, it was shown that the TEE drops a half at $T=0^+$, remains constant when $0<T<T_c$, then drops to zero at a critical temperature $T=T_c$, see Fig.~\ref{pic-piecewise}(b). Although $\mathrm{TEE}>0$ at low temperature, the authors in Ref.~\onlinecite{Chamon3d} argued that it is just a classical memory. 

It is therefore a natural question to consider the finite-temperature TEE for fracton models. In this paper, instead of solving a specific model, we analyze this problem for general Calderbank-Shor-Steane (CSS) codes~\cite{CSS1,CSS2}, which include many familiar examples for ordinary topological orders and fractonic topological orders. 

In Sec.~\ref{sec-TEE}, we prove that for any topological-ordered CSS codes (ordinary or fractonic), for any definition of the TEE (as will be explained in the main text, the definition of TEE has some ambiguities so one need to make a choice), the TEE is a piecewise constant function of the temperature. Possible discontinuities happen only at phase transition temperatures. Since it is relatively easy to calculate the TEE at zero temperature \cite{MahanTEE,HehuanTEE}, and at high enough temperature the system should be disordered with TEE=0 (for example, at infinite temperature, all local degrees of freedom decouple, no entanglement at all), this theorem is enough for us to determine the TEE at all temperature in some cases (for example, if the model has only one phase transition). As we will see, even if we cannot determine the TEE to its precise value (it depends on a choice anyway), it provides us enough information in many models.

The problem is now reduced to the phase structures. In Sec.~\ref{sec-example} we derive the partition function for four representative models, namely, 2D/3D toric code, X-cube model, and Haah's code. Notably, in most cases, we only need an inequality instead of brute force calculations. The precise value of the TEE can also be determined in these examples. In all cases, there is a phase transition at $T=0$ and a corresponding drop in the TEE. In Sec.~\ref{sec-general}, based on the low temperature expansion and the existence of fractal generators, we argue that any CSS code in 2D and 3D with topological orders has a phase transition at $T=0$. This indicates the break down of topological orders at finite temperature.

\section{Finite-temperature topological entanglement entropy}\label{sec-TEE}
In this section, we review some necessary calculations of finite-temperature topological entanglement entropy for CSS codes.

\subsection{CSS codes}
In this paper, we will consider toric-code-like stabilizer codes in $D$ space dimensions, called CSS codes \cite{CSS1,CSS2}, named after three authors of the references. We will always assume translational invariance. So without loss of generality, our model lives on $\zz^D$ lattices.

For each point in $\zz^D$ (correspond to a unit cell), we put $q$ qubits (bosonic spin-$1/2$) on it ($q\geq 1, q\in\zz$). Consider Hamiltonians of the following form:
\begin{equation}\label{eq-CSS}
H=-\lambda_A\sum_i{A_i}-\lambda_B\sum_j{B_j},
\end{equation}
where $A_i/B_j$ means some local products of Pauli $X/Z$ operators around position $i/j$ ($i,j\in\zz^D$). We will always assume $[A_i,B_j]=0$, so our models are stabilizer codes~\cite{NielsenChuang}. Note that we can have more than one type of products $A$ and $B$. 

We will mainly consider the followings representative examples.
\begin{itemize}
\item  \textbf{2D toric code} \cite{Kitaev-TQC}. Qubits live on the links, so $q=2$. Here, $A$ is the star operator, defined as the product of 4 Pauli $X$s on the 4 links connected to a point. $B$ is the plaquette operator, defined as the product of 4 Pauli $Z$s on the 4 links around a plaquette (2-cell).
\item \textbf{3D toric code} \cite{3dtoric}. Qubits live on the links, so $q=3$. Here, $A/B$ is still the star/plaquette operator as before. However, we have three plaquette operators since in this lattice there are three different 2-cells ($xy,yz,zx$).
\item \textbf{X-cube model} \cite{Xcube1st,FSduality}. This is a 3D model with qubits live on the links, so $q=3$. $A$ is the star operator, defined as the product of 4 Pauli $X$s on the 4 links in a 2-dimensional plane. So we have three different types of star operators $A_{xy},A_{yz},A_{zx}$, although there is a local relation $A_{xy}A_{yz}A_{zx}=1$. $B$ is the cubic operator, defined as the product of 12 Pauli $Z$s on the 12 links around a cube (3-cell). It is an example of type-I fractons \cite{FSduality}.
\item \textbf{Haah's code} \cite{Haahcode}. This is a 3D model with 2 qubits on each point, $q=2$. Here $A$ and $B$ are defined as in the following figure. It is an example of type-II fractons.
\begin{figure}[h!]
	\centering
	\includegraphics[width=0.7\columnwidth]{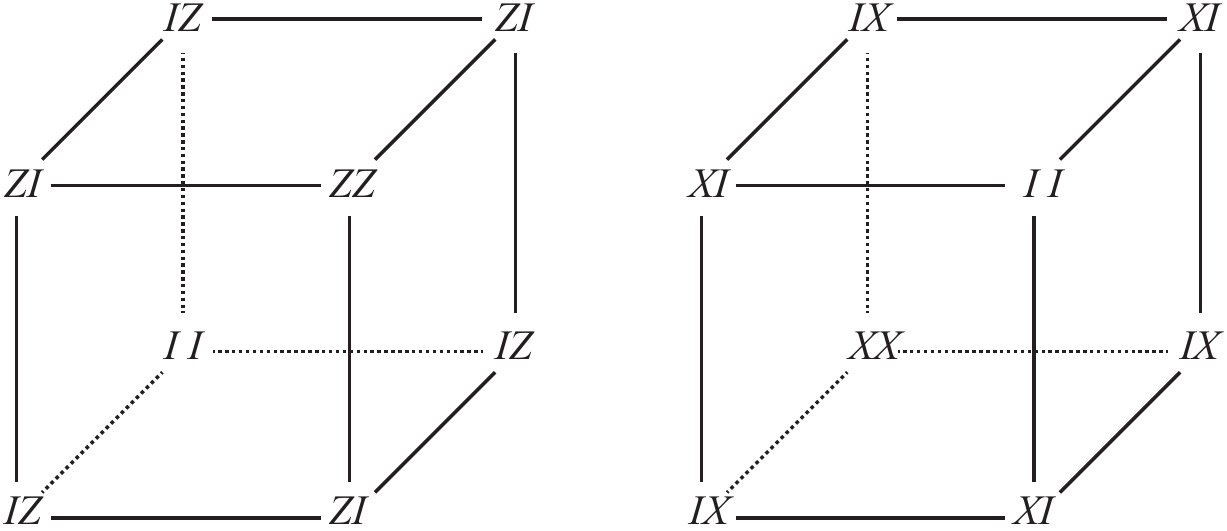}	
\end{figure}
\end{itemize}

\subsection{topological entanglement entropy}
Let us consider a bi-partition of the systems as $C\cup D$. If the whole system is in the (maybe mixed) state $\rho$, then the entanglement entropy on subsystem $C$ of a partition is defined by
\begin{equation}\label{eq-entropy}
S_C=-\Tr(\rho_C\ln\rho_C),
\end{equation} 
where $\rho_C=\Tr_D(\rho)$ is the reduced density matrix. If $\rho$ is pure, $S_C=S_D$. However since we will consider finite temperature, $\rho\propto\exp(-\beta H)$, we do not have such equation.

\begin{figure}[htp]
	\centering
	\includegraphics[width=\columnwidth]{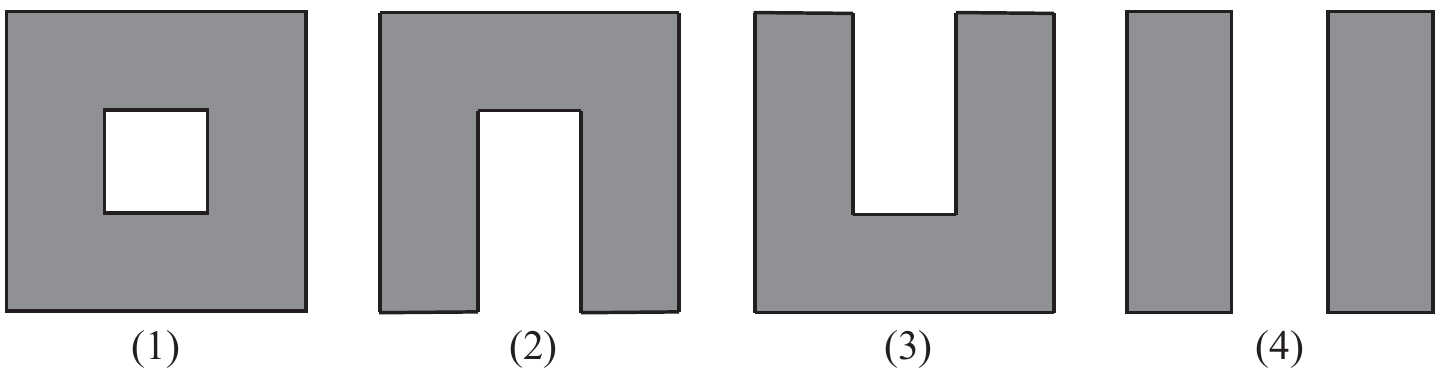}
	\caption{A bi-partition scheme to define the topological entanglement entropy in 2D.}		\label{pic-partition}
\end{figure}

In order to define topological entanglement entropy, following \cite{Kitaev-tee,Wen-tee}, we need a combination of different bi-partitions to cancel the leading contribution(s). For example, in two dimensions, we can use the bi-partitions shown in Fig.~\ref{pic-partition}. These bi-partitions are designed such that the volume contribution and the area contribution are cancelled exactly in the following combination:
\begin{equation}\label{eq-teedef}
\TEE=\lim_{r,R\to\infty} (S_1-S_2-S_3+S_4),
\end{equation}
where $S_i$ is the entanglement entropy for corresponding bi-partition in the figure, $r$ and $R$ are the size of the inner square and outer square.

One can definitely use different bi-partition schemes. Logistically speaking, different schemes give different results for the topological entanglement entropy. This is indeed what  happens in some fracton models \cite{MahanTEE}. Our result will be valid for all possible bi-partition schemes.

\subsection{finite temperature calculation}
The definition of entanglement entropy \refeq{eq-entropy} involving the logarithm seems complicated at first. Remarkably, for CSS codes, one can perform the calculation to a large extent and get a quite compact result. In this subsection, we review the necessary results in Refs.~\onlinecite{Chamon2d,Chamon3d,Hamma}.

In \refeq{eq-CSS}, denote $H_A=-\sum A$ and $H_B=-\sum B$ so that $H=\lambda_AH_A+\lambda_BH_B$. We work in $Z$-basis, i.e., use $\{\ket{f}\}$ as a basis of the (many-body) Hilbert space where $f$ is a configuration of all spins in $Z$-basis. We then have:
\begin{equation}
e^{-\beta H}=\sum_{f,f'}\ket{f'}\bra{f'}e^{-\beta H}\ket{f}\bra{f}=\sum_{f,f'}\bra{f'}e^{-\beta H}\ket{f}\ket{f'}\bra{f}.
\end{equation}
Since $H_B$ is diagonal in $Z$-basis, we have:
\begin{equation}
\bra{f'}e^{-\beta H}\ket{f}=\bra{f'}e^{-\beta\la H_A}\ket{f}e^{-\beta\lb H_B(f)},
\end{equation}
where $H_B(f)=\bra{f}H_B\ket{f}$. Then we need two observations: 
\begin{itemize}
	\item $\bra{f'}e^{-\beta\la H_A}\ket{f}\neq 0$ only if $\ket{f'}$ can be obtained by acting some $A$ operators (which flip some spins) on $\ket{f}$. Define $G$ be the group generated by all possible products of $A$ operators. It is an Abelian group in which all elements square to 1. Sum over $(f,f')$ is then equivalent to sum over $(f,gf)$ where $f'=gf$ (flip $\ket{f}$ by $g \in G$). 
    \item $\bra{gf}e^{-\beta\la H_A}\ket{f}$ is independent of $f$ since $\bra{ghf}e^{-\beta\la H_A}\ket{hf}=\bra{f}hge^{-\beta\la H_A}h\ket{f}$ and $hge^{-\beta\la H_A}h=ge^{-\beta\la H_A}$ for $\forall h$ where $h$ is a product some Pauli $X$ operators that takes $\ket{f}$ to another configuration. Let's denote it by $p(g,\la)$.
\end{itemize}
Therefore,
\begin{equation}\label{eq-factorizeZ}
\begin{aligned}
&e^{-\beta H}=\sum_{g,f}p(g,\la)e^{-\beta\lb H_B(f)}g\ket{f}\bra{f},\\
&Z=\Tr e^{-\beta H}=p(1,\la)\sum_{f}e^{-\beta\lb H_B(f)}=p(1,\la)Z_B,
\end{aligned}
\end{equation}
where $Z_B$ is exactly the partition function in the case of $\la=0$, i.e., throw away $H_A$ terms.

Now we can take the partial trace. Divide the system into two subsystems $C$ and $D$, then we can factorize $g=g_C\otimes g_D$ and $\ket{f}=\ket{f_C}\otimes\ket{f_D}$. Therefore,
\begin{equation}\label{eq-rho1}
\begin{aligned}
\Tr_D g\ket{f}\bra{f}&=\Tr_D \ket{g_Cf_C}\otimes\ket{g_Df_D}\bra{f_D}\otimes\bra{f_C}\\
&=\delta_D(g_D)\ket{g_Cf_C}\bra{f_C},\\
\rho_C=\frac{\Tr_D e^{-\beta H}}{Z}&=\sum\limits_{g\in G_C,f}\frac{p(g,\la)}{p(1,\la)}\frac{e^{-\beta\lb H_B(f)}}{Z_B}\ket{gf_C}\bra{f_C},
\end{aligned}
\end{equation}
where $\delta_D(g_D)$ is the delta function to impose $g_D=1_D$, $G_C=\{g|g_D=1_D\}$ is a subgroup of $G$ such that all elements act on subsystem $D$ trivially.

To obtain the entanglement entropy, we will use the replica trick:
\begin{equation}
S_C=-\Tr(\rho_C\ln\rho_C)=-\lim_{n\to 1}\partial_n\Tr{\rho_C^n}.\label{eq-replica}
\end{equation}
Using \eqref{eq-rho1} we get:
\begin{equation}
\begin{aligned}
\Tr{\rho_C^n}=\sum_{g_i\in G_C}\sum_{f_i}\left(\prod_{i=1}^n\frac{p(g_i,\la)}{p(1,\la)}\right)\left(\prod_{i=1}^n\frac{e^{-\beta\lb H_B(f_i)}}{Z_B}\right)\\
\braket{f_n|g_1f_1}_C\braket{f_1|g_2f_2}_C\cdots\braket{f_{n-1}|g_nf_n}_C,
\end{aligned}
\end{equation}
where $f_i$ still means configurations on the full system $C\cup D$ and $\braket{f|f'}_C$ means $\braket{f_C|f'_C}$, the inner product on the Hilbert space for subsystem $C$.

Note that
\begin{equation}
\begin{aligned}
&\braket{f_n|g_1f_1}_C\braket{f_1|g_2f_2}_C\cdots\braket{f_{n-1}|g_nf_n}_C\\
=&\begin{cases}
1, & \prod_{i=1}^ng_i=1_C \text{~and}\\
& f_k=g_k\cdots g_2f_1 \text{~on $C$ for~} k=2,\cdots,n\\
0, & \text{otherwise}
\end{cases}.
\end{aligned}
\end{equation}
and that $H_B(f)=H_B(gf)$ since $g=\prod A$ and $[A,B]=0$, we get (after a transformation $f_k\to g_k\cdots g_2f_1,~k=2,\cdots n$)
\begin{equation}
\begin{aligned}
\Tr{\rho_C^n}=&\sum_{g_i\in G_C}\left(\prod_{i=1}^n\frac{p(g_i,\la)}{p(1,\la)}\right)\delta_C(\prod_{i=1}^ng_i)\\
\times&
\sum_{f_i}\left(\prod_{i=1}^n\frac{e^{-\beta\lb H_B(f_i)}}{Z_B}\right)\delta_C(f_1,f_2,\cdots,f_n),
\end{aligned}
\end{equation}
where $\delta_C(\prod_{i=1}^ng_i)$ imposes $\prod_{i=1}^ng_i=1_C$ and $\delta_C(f_1,f_2,\cdots,f_n)$ imposes $f_1=f_2=\cdots =f_n$ on $C$.

Importantly, two terms in the product only depends on $\la$ and $\lb$ respectively. Therefore we have the following factorization\footnote{Assume $f(x,y)=g(x)h(y)$ then $f(x,0)=g(x)h(0)$ and $f(0,y)=g(0)h(y)$. Therefore $f(x,y)=f(x,0)f(0,y)/h(0)g(0)$. Let $\la=\lb=0$, there is no dynamics, $\rho_C=\frac{\mathbbm{1}}{2^{|C|}}$, so the coefficient is determined.}:
\begin{equation}\label{eq-factorize}
\Tr{\rho_C^n}(\la,\lb)=2^{n|C|}\Tr{\rho_C^n}(\la,0)\cdot\Tr{\rho_C^n}(0,\lb),
\end{equation}
where\footnote{The ``$\propto$" is actually ``=" since $p(g,\la=0)=\bra{f}g\ket{f}=\delta(g)$. }
\begin{equation}
\Tr{\rho_C^n}(0,\lb)\propto \sum_{f_i}\left(\prod_{i=1}^n\frac{e^{-\beta\lb H_B(f_i)}}{Z_B}\right)\delta_C(f_1,f_2,\cdots,f_n).
\end{equation} 

Now let's calculate the entanglement entropy. From \refeq{eq-replica} and \refeq{eq-factorize} it's easy to show the (topological) entanglement entropy is the summation of two independent contribution:
\begin{equation}
\TEE(\la,\lb)=\TEE(\la,0)+\TEE(0,\lb).
\end{equation}
From now on we will just set $\la=0$ and hide subscripts $B$. We apply a decomposition $f=f_C\otimes e$ where $f_C/d$ is a configuration on $C/D$ respectively. Then $\sum_{f_1,\cdots,f_n}\delta_C(f_1,f_2,\cdots,f_n)=\sum_{f_C}\sum_{d_1,\cdots,d_n}$. Therefore,
\begin{equation}
\begin{aligned}
\Tr{\rho_C^n}(0,\lambda)
&=\sum_{f_C}\sum_{d_1,\cdots,d_n}\prod_{i=1}^n\frac{e^{-\beta\lambda H(f_Cd_i)}}{Z}\\
&=\sum_{f_C}\left(\frac{\sum_{d}e^{-\beta\lambda H(f_Cd)}}{Z}\right)^n\defeq \sum_{f_C}q_C(f)^n.
\end{aligned}
\end{equation}
Note that $q_C(f)$ is a probability distribution on $F_C$: $\sum_{f_C}q_C(f)=1$, hence
\begin{equation}
\begin{aligned}
S_C&=-\sum_{f_C}q_C(f)\ln q_C(f)\\
&=-\sum_{f}\frac{e^{-\beta\lambda H(f)}}{Z}\ln [q_C(f)Z]+\ln Z.
\end{aligned}
\end{equation}
The meaning of the first line is obvious: set $\la=0$, the classical distribution over possible classical configurations $f$ induces a classical distribution on subsystem C, $q_C(f)$, whose entropy is exactly the entanglement entropy of $B$ sector.

The second line is more convenient for topological entanglement entropy calculation, since the summation $\sum_{f}$ is over the (fixed) whole system instead of the subsystem $C$: recall we need to vary the partition $C\cup D$ as in \refeq{eq-teedef}. We finally get:
\begin{equation}\label{eq-teefinal}
\TEE=-\sum_{f}\frac{e^{-\beta\lambda H(f)}}{Z}\sumtop\ln [q_C(f)Z],
\end{equation}
where $\sumtop$ is the linear combination required by \refeq{eq-teedef}.

\section{piecewise constancy of entanglement entropy}
\newcommand{\pb}{\partial_\beta}
To prove the piecewise constancy of $\TEE$, we first show $\sumtop\ln [q_C(f)Z]$ is piecewise constant as a function of inverse temperature $\beta$. We will consider its dependence on $f$ later. 

Let us consider $\pb \ln[q_C(f)Z]$. We have:
\begin{align}
\pb \ln[q_C(f)Z]=\frac{-\sum_{d}\lambda H(f_Cd) e^{-\beta\lambda H(f_Cd)}}{\sum_{d}e^{-\beta\lambda H(f_Cd)}}=\lambda \sum_i\braket{B_i}_f,
\end{align}
where $\braket{H}_f$ means the average of $H$ under the condition of fixing $f_C$.

We will consider $\beta$ such that the system has no long range correlation of $B$ operators, i.e, all correlation functions decay exponentially. For our model, this implies no long range correlation for all operators, due to Elizur's theorem\footnote{This classical spin model has local symmetries given by $A$ (and only $A$, since the original CSS model is complete: no other independent stabilizers). According to Elizur's theorem, operators with nonzero expectation value must be local-symmetry-invariant, which must be products of $B$ and (global) logical operators.}. 
We assume that this condition is violated only for some discrete $\beta$ (i.e., critical temperature). In other words, we assume there is no extended ordered phases\footnote{One may worry about some non-critical phases with long range order, like Ising model in low temperature. However, this will not happen here because there is no symmetry breaking where $B$ serves as an order parameter, due to local indistinguishability. Moreover, even for Ising model in the low temperature phase, the bond-bond correlation (which arises as the $B$ operator in the Hamiltonian) is still short ranged \cite{Wegner}.} or critical phases. In this case, $\braket{B_i}_f$ should approach an $f$-independent value as $i$ becomes far from $C$. This is because we only effectively fix the configuration inside $C$, $f_C$, when calculating $\braket{B_i}_f$, which should have exponentially vanishing effects on $B_i$ if $i$ is far away from $C$.

For each partition $C\cup D$, denote $\overline{C}=C+$``a shell with thickness $O(\ln N)$" where $N$ is the total number of qubits. The actual coefficient of $\ln N$ is not important, as long as it is the same for all bi-partitions and is big enough to ensure the following $o(\frac{1}{N})$. For $i\notin\overline{C}$, one has
\begin{equation}
    \braket{B_i}_f=\braket{B_\infty}+o(\frac{1}{N})
\end{equation}
since $i$ is $O(\ln N)$ far away from $C$; for $i\in C$, $\braket B_i$ is fully determined by $f$; for $i\in \overline{C}-C$, $\braket{B_i}_f$ is determined by $f$ near $i$ (radius$\sim O(\ln N)$) up to error $o(\frac{1}{N})$. Note that we require $\ln N\ll L $ while still $L\ll N$, where $L$ is the linear size of the partition $C$, so that each $\braket{B_i}_f$ depends only on the small region near the partition boundaries.

Before going on, we discuss two special cases where this can be seen more clearly. In both cases, the $o(\frac{1}{N})$ is exactly zero.
\begin{itemize}
	\item If there is are string operators connecting $B$ excitations and $\braket{B_\infty}_f$ is independent on $f_C$. In this case, consider $\braket{B_i-B_j}_f$, we have:
	\begin{equation}
	\begin{aligned}
	&\braket{B_i-B_j}_f\\
	=&\frac{2}{q_C(f)Z}\left[\sum_{\substack{B_i=1\\B_j=-1}}e^{-\beta\lambda H(f_Cd)}-\sum_{\substack{B_i=-1\\B_j=1}}e^{-\beta\lambda H(f_Cd)}\right].
	\end{aligned}
	\end{equation}
	Use a string operator that flips the value of $B_i, B_j$, we get a one-to-one correspondence between terms in two summations and therefore $\braket{B_i}_f=\braket{B_j}_f$. If we have plenty of string operators as in the case of 2D toric code, we see that $\braket{B_i}_f=\braket{B_\infty}_f$ exactly for all $i$. Moreover, it's enough to assume that one can connect each $i$ with a $j$ (may depend on $i$) arbitrarily far away, as in the case of the star sector in the X-cube model.

	\item If $B$ excitations are ``almost free" in the sense of Sec.~\ref{sec-example}. Examples are 2D toric code, X-cube model, Haah's code, and stars in 3D toric code. In this case, $B_i$ behaves like independent spins as we will see in Sec.~\ref{sec-example}. $\braket{B_i}_f$ is therefore independent of $i$ and $f$.
\end{itemize}

Now take the topological combination of partitions. Since the interiors and boundaries of these $C$ are designed to cancel, we have:
\begin{align}\label{eq-Tdenpendce}
\begin{autobreak}
\pb \sumtop\ln[q_C(f)Z]
=\lambda \sum_i\sumtop\braket{B_i}_f
\lesssim No(\frac{1}{N})+L^{D-1}\ln N o(\frac{1}{N})
=o(1),
\end{autobreak}
\end{align}
which vanishes in the thermodynamics limit. 

However, $\ln [q_C(f)Z]$ may still depend on $f$. 
To proceed, we rewrite \refeq{eq-teefinal} as
\begin{equation}\label{eq-teefinal2}
\TEE=-\sum_{f}\frac{e^{-\beta\lambda H(f)}}{Z}\sumtop\overline{\ln [q_C(f)Z]},
\end{equation}
where $\overline{\ln [q_C(f)Z]}$ means average over all possible $f'$ with the same number of $B$ excitations as $f$. We can do this because $H(f)$ only depends on this number.

Now consider $f$ and $1$ (the configurations where all spins are upward), then:
\begin{equation}\label{eq-fdependence}
\ln [q_C(f)Z]-\ln [q_C(1)Z]=\ln\frac{\sum_{d}e^{-\beta\lambda H(f_{C}d)}}{\sum_{d}e^{-\beta\lambda H(1_{C}d)}}.
\end{equation} 
The denominator is a restricted partition function of our model (fix all spins in $C$ to be up and only varies the spin configuration $d$ on $D$). The numerator can be regarded as a partition function of a different model: one still fixes all spins in $C$ upward and only varies $d$ on $D$, however, the sign of some stabilizers $B$ in $H$ is flipped (from negative to positive) if this $B$ itself is excited for configuration $f$. In other words, the new model is equal to the old model with ``magnetic dislocations" \cite{Wegner} to flip some couplings. Therefore, the quotient is $\braket{\prod_f M_B}_{1C}$ where $M_B$ is the dislocation operator \cite{Wegner} for a $B$ term:
\begin{equation}
M_B=e^{-2\beta\lambda B}=\cosh 2\beta\lambda-\sinh 2\beta\lambda B,
\end{equation}
and $\braket{~}_{1C}$ is the expectation value under the condition that all spins are upward in $C$. Therefore,
\begin{equation}
\overline{\ln [q_C(f)Z]}-\overline{\ln [q_C(1)Z]}=\overline{\ln {\braket{\prod M_B}_{1C}}}.
\end{equation}
The last average can be regarded as average over dislocation configurations with total number $m$ of dislocations operators $M_B$ fixed. 

Importantly, being a linear function of $B$, the dislocation operators are also short range correlated. Therefore, for a fixed $m$, most configurations are where all $M_B$ are far from $C$. The average in the thermodynamics limit will only see those typical configurations and will be independent of $C$. Therefore,
\begin{equation}
\sumtop\overline{\ln [q(f)Z]}=\sumtop\overline{\ln [q(1)Z]}=\sumtop{\ln [q(1)Z]},
\end{equation}
which is then independent of $f$. Plug it into \refeq{eq-teefinal2}, we get:
\begin{equation}
    \TEE=-\sumtop{\ln [q(1)Z]},
\end{equation}
and is therefore a piecewise constant function of the temperature.

\section{Phase transition: examples}\label{sec-example}
According to the piecewise constancy, it is important to consider the phase structure of a CSS model. In this section, we consider phase transitions in four representative models: 2D and 3D toric codes, the X-cube model, and Haah's code. We will prove the existence of a zero-temperature phase transition in these models. We will actually prove a stronger statement: to the extent of thermodynamics, these model contains a \textbf{``free sector"} in the sense that it behaves like independent spins, thus a phase transition happens at zero temperature.

First, a general remark. From \refeq{eq-factorizeZ} we know the partition function factorized as
\begin{equation}
Z(\beta)\propto Z_A(\beta)Z_B(\beta),
\end{equation}  
where $Z_{A/B}(\beta)$ is the partition function of a classical spin model defined by keeping only the $A/B$ part of the original Hamitonian \eqref{eq-CSS}. The problem of phase transitions of the original quantum model is reduced to two classical sectors.

\subsection{2D toric code}
Let's consider the 2D toric code to illustrate our idea. Due to the electric-magnetic duality in 2D toric code, we only need to consider $Z_B(\beta)$ (set $\lb=1$):
\begin{equation}
Z_B(\beta)=\sum_{\{s_i\}} e^{-\beta H\{s\}}=\sum_{\{s_i\}}\prod_ie^{\beta B_i}.
\end{equation}  
Since $B_i=\pm 1$, we want to use $B_i$ as elementary degree of freedoms (classical spins). The only constraint is $\prod B_i=1$. More precisely, it's not hard to see 
\begin{equation}\label{eq-spinasdof}
\sum_{s_i}=2^{N+1}\sum_{\{B_i\}|\prod B_i=1}=2^{N+1}\sum_{\{B_i\}}\frac{1+\prod_i B_i}{2},
\end{equation}
where $2^{N+1}$ is because the number of physical spins is $2N$ and the number of independent $B$ operators is $N-1$. Thus,
\begin{equation}
\begin{aligned}
Z_B(\beta)&=2^{N+1}\sum_{\{B_i\}}\frac{1+\prod_i B_i}{2}\prod_ie^{\beta B_i}\\
&=2^{N}\prod_i(\sum_{B_i}e^{\beta B_i})+2^{N}\prod_i(\sum_{B_i}B_ie^{\beta B_i})\\
&=2^{2N}(\cosh^N\beta+\sinh^N\beta)\\
&=2^{2N}\cosh^N\beta(1+\tanh^N\beta).
\end{aligned}
\end{equation}
What determined the phase transition is the free energy per volume in the thermodynamics limit:
\begin{equation}\label{eq-irr}
f=\lim_{N\to\infty}-\frac{\ln Z_B(\beta)}{\beta N}=-\frac{1}{\beta}(2\ln 2+\ln\cosh\beta).
\end{equation}
So the partition function is a smooth function for $T>0$ and has a singularity at $T=0$. This means a phase transition at zero temperature and no phase transition at any finite temperature.

Intuitively, what happened is: if we regard $B_i$ as elementary spins, they are almost free except the constraint $\prod B_i=1$. However, \textbf{only one global constraint is irrelevant in the thermodynamic limit}: as we see in the calculation, the only effect is to multiply a factor $(1+\tanh^N\beta)$, which contributes 0 to $f$ anyway. 

In conclusion, each sector of 2D toric code behaves like independent spins in the thermodynamics limit. Therefore, the 2D toric code has only one phase transition, which is at $T=0$. According the piecewise constancy, we reproduce the result in Ref.~\onlinecite{Chamon2d} without heavy calculations: the topological entanglement entropy $\TEE=0$ for all $T>0$.

\subsection{Haah's code, X-cube model, and 3D toric code}
The situation is only slightly more complicated in Haah's code. In this case, the number of constraints $k$ is at most $O(L)$ ($L$ is the linear size, $N=L^3$), since it's essentially the ground state degeneracy. The constraints enter the partition function through
\begin{equation}
\prod_k{(1+\prod B_i)}=1+\prod B+\sideset{}{'}\prod B+\cdots,
\end{equation}
where $\prod_k$ means product of all $k$ constraints, and other $\prod$ is a product of some (do not need the details) $B$ operators. It gives a factor $(1+\tanh^a\beta+\tanh^b\beta+\cdots)$
where $a,b\in[0,N]$ (unimportant), which contributes 0 to $f$ as in \refeq{eq-irr}, since
\begin{equation}
\frac{1}{N}\ln(1+\tanh^a\beta+\tanh^b\beta+\dots)<\frac{\ln 2^k}{N}\to 0.
\end{equation}
Therefore, phase structure and the behaviour of $\TEE$ of the Haah's code is the same as 2D toric code: only one phase transition which is at $T=0$ and $\TEE=0$ for all $T>0$, see Fig.~\ref{pic-piecewise}(a).

In the case of the X-cube model, cubes and stars are not equivalent. For the cubic interaction, $k=O(L)$ (the product of cubes along each 2-dimensional plane is 1), the analysis of Haah's code still applies here. For the star interaction, we have local constraints $A_{xy}A_{yz}A_{zx}=1$ besides $O(L)$ global constraints. These constraints can be \textbf{eliminated}. Denote $A_{xy}=P, A_{yz}=Q$, then
\begin{equation}\label{eq-partitionX}
Z_A(\beta)\propto\sideset{}{'}\sum_{P,Q}e^{\beta\sum(P+Q+PQ)}
\end{equation}
where $\sum'$ means summation system of independent subsystems, where each site has two classical spins with $H=-\sum(P+Q+PQ)$.

Let's prove that the global constrains are indeed irrelevant. Similar as before,
\begin{equation}
\sideset{}{'}\sum=\sum\prod_k\frac{(1+\prod P\prod Q)}{2}\propto\sum( 1+\prod P\prod Q+\dots).
\end{equation} 
Plug into \refeq{eq-partitionX}, we get
\begin{equation}
\sum \prod e^{\beta(P+Q+PQ)}+\sum\prod P\prod Q \prod e^{\beta(P+Q+PQ)}+\dots.
\end{equation}
The first term equals to $(e^{3\beta}+3e^{-\beta})^N$. Each following term is a product of terms of two classes, depending on whether $P$ or $Q$ appears in corresponding $\prod P\prod Q$:
\begin{equation}
\begin{cases}
e^{3\beta}+3e^{-\beta}, & \text{no~} P,Q \text{~appears}\\
e^{3\beta}-e^{-\beta}, & \text{otherwise}
\end{cases}.
\end{equation}
Therefore, 
\begin{equation}
Z(\beta)\propto (e^{3\beta}+3e^{-\beta})^N\left(1+\left(\frac{1-e^{-4\beta}}{1+3e^{-4\beta}}\right)^a+\dots\right)
\end{equation}
and then similar arguments go through.

Therefore, phase structure (see also Ref.~\onlinecite{Xcube-others}) and the behavior of $\TEE$ of X-cube model is also similar to 2D toric code, see Fig.~\ref{pic-piecewise}(a).

For 3D toric code, the physics is different. Here, the star interaction is still almost free, with only one global constraint given by the product of all stars. The plaquette sector has local constraints: products of 6 plaquettes around a cube is 1, which however cannot be eliminated as in the case of X-cube stars. In \cite{Chamon3d,Wegner}, it is shown that the plaquette part in 3D toric code is dual to the 3D Ising model, thus the phase transition is at finite temperature. Therefore, 3D toric code has two phase transitions, one at $T=0$, one at $T=T_c\neq 0$. The $\TEE$ is a piecewise constant function with two drops as shown in Ref.~\onlinecite{Chamon3d}.

In conclusion, for four models considered here, there is a zero-temperature phase transition due to the existence of a ``free sector". In 2D toric code, X-cube model and Haah's code, there is no other phase transitions at finite temperature while in 3D toric code there is a phase transition at finite temperature.

It is natural to ask the phase structures of general CSS codes. We will show in Appendix \ref{app-counter} that in general we do not have a ``free part" or even a part where local constraints can be eliminated (like the star operators in the X-cube model). However, in the next section, we argue that a zero-temperature phase transition exists as long as the system has topological order.

\section{Phase transition: general arguments}\label{sec-general}
In Sec.~\ref{sec-example}, the strategy to show the existence of zero-temperature phase transition is ``global": we calculate the partition function for all temperature. In this section, we will pursue a ``local" approach: to study the model near $T=0$, by looking at the low temperature expansion.

As a warm-up, consider the low temperature expansion \cite{Pathria} of the 2D Ising model. We set the coupling to be 1, so $H=-\sum{s_is_j}$. The partition function expanded near the ground states is
\begin{equation}
\begin{aligned}
Z(\beta)&=\sum_{\{s\}} e^{-\beta H}=e^{-\beta E_0}\sum_{\{s\}} e^{-\beta (H-E_0)}\\
&=e^{-\beta E_0}\sum_{\{s\}} x^{M(\{s\})},
\end{aligned}
\end{equation}
where $E_0=-2N$ is the ground state energy (all bonds $b_{ij}=s_is_j=1$), 
\begin{equation}
    x=e^{-2\beta},
\end{equation} 
and $M(\{s\})$ is the number of flipped bonds ($b_{ij}=-1$). Since the products of all bonds must be 1, the number flipped bonds is even.

$M=0$: ground states, two configurations: all spin up and all spin down.

$M=2$: This is impossible.

$M=4$: The only possibility is to flip a spin relative to a ground state, so that we get a ``star" configuration of bond excitations. There are $2N$ ways to do it.

Thus, 
\begin{equation}
Z(\beta)=2e^{-\beta E_0}(1+Nx^4+\dots).
\end{equation}
If we calculate the free energy, we will find:
\begin{equation}
f=-\frac{\ln Z_B(\beta)}{\beta N}=-\frac{1}{\beta N}(\ln 2+2\beta N+ N x^4+\dots).
\end{equation}
In the thermodynamic limit $N\to\infty$, there is no problem at least up to order $x^4$ since the coefficient of it is $O(N)$.

In contrast, consider the low temperature expansion of the plaquette sector of 2D toric code. We still have:
\begin{equation}
Z(\beta)=e^{-\beta E_0}\sum_s x^{M(\{s\})},
\end{equation}
where $E_0=-N$, $M(\{s\})$ is the number of flipped plaquettes. The leading order will be $x^2$ since we can have two plaquette excitations. However, there are $\binom{N}{2}$ plaquette configurations with two excitations since any string operator will produce an excitation at each end. Therefore,
\begin{equation}
    Z(\beta)=2^{N+1}e^{-\beta E_0}(1+\binom{N}{2}x^2+\cdots),
\end{equation}
where $2^{N+1}$ is due to ```gauge transformations" as in \refeq{eq-spinasdof}, and:
\begin{equation}\label{eq-breakdown}
f=-\frac{\ln Z_B(\beta)}{\beta N}=-\frac{1}{\beta}\left(\text{const}+\beta+ \frac{\binom{N}{2}}{N}x^2+\dots\right),
\end{equation}
which has no thermodynamic limit due to the coefficient $\binom{N}{2}$. The break down of low energy expansion indicates there is a zero-temperature phase transition in the 2D toric code.

\begin{figure}[h]
	\includegraphics[width=\columnwidth]{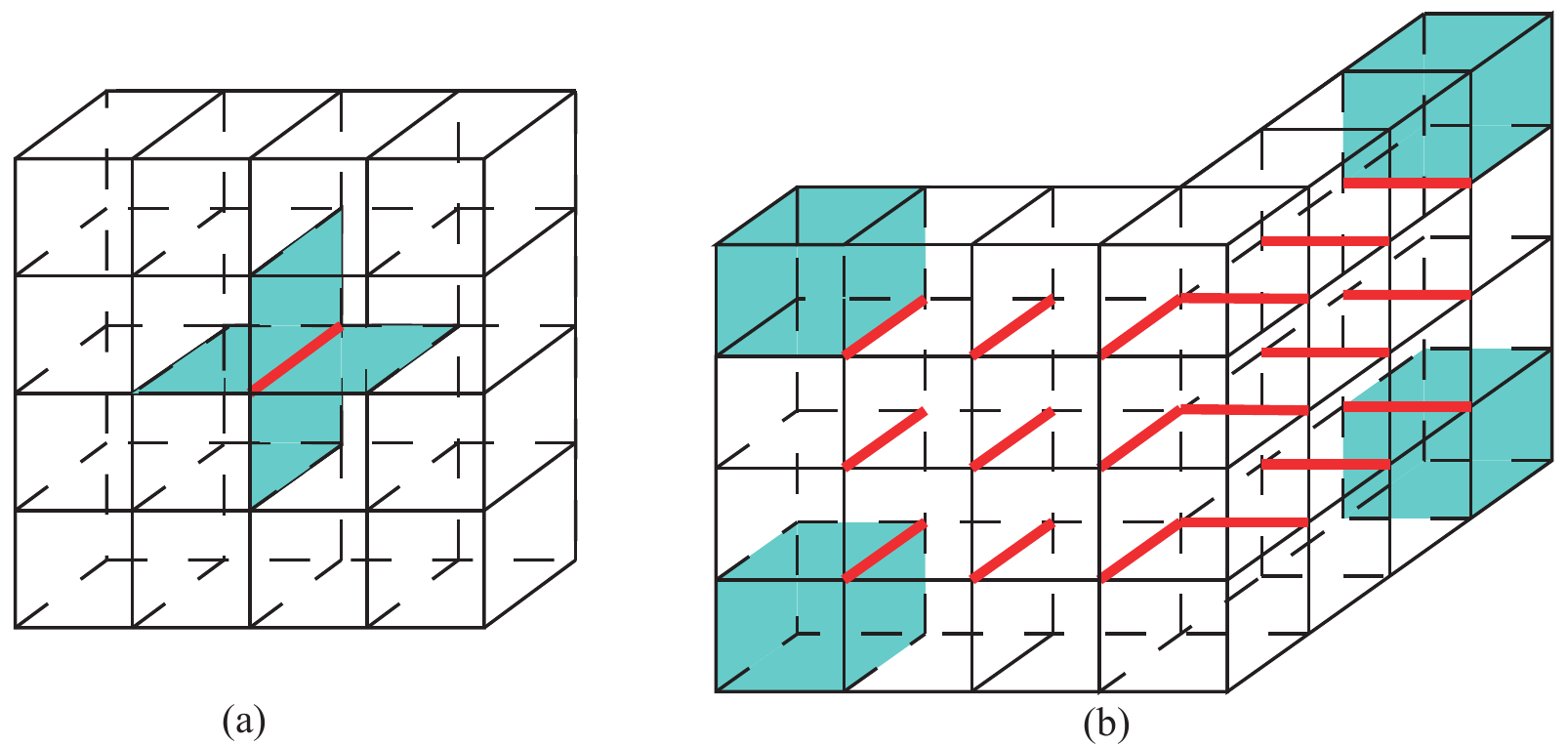}
	\caption{Leading terms in the low temperature expansion for 3D toric code and X-cube model. (a) Plaquette sector in 3D toric code, minimal excitation is a windmill shaped configuration by flipping one spin (on the red link). (b) Cubic sector in X-cube model, minimal excitation is caused by a bended membrane operator (shown red).}\label{pic-expansion}
\end{figure}

It's not hard to work out the low temperature expansion for 3D toric code and the X-cube model at the leading order.
\begin{itemize}
\item \textbf{3D toric code}. The minimal number of star excitations is 2, generated by a string operator at its ends. So similar to Eq.~\eqref{eq-breakdown}, there is a $\binom{N}{2}x^2$ term breaking down the expansion, which indicates a zero-temperature phase transition in this sector. 

The minimal number of plaquette excitations is 4, given by a ``windmill" configuration, see Fig.~\ref{pic-expansion}(a). The coefficient of $x^4$ is $2N$, so no problem in the expansion at least to the leading order, which indicates no zero-temperature phase transition in this sector.

\item \textbf{X-cube model}. The minimal number of cubic excitations is 4, produced by a bended membrane operator, see Fig.~\ref{pic-expansion}(b). It's not hard to see the coefficient of $x^4$ is $O(L^2\!\times\! L^2\!\times\! L\!\times\! L)=O(N^2)$, indicating a zero-temperature phase transition of the cubic sector. 

For the star excitations, the minimal number of excitations is 4, produced by a straight string operator (2 stars at each end). The coefficient of $x^4$ is $O(L^2\!\times\! L\!\times\! L)=O(N^{4/3})$, indicating a zero-temperature phase transition of the star sector.
\end{itemize}

In these models, the results based on the low temperature expansion agree with those from Sec.~\ref{sec-example}.

\begin{figure}
	\centering
	\includegraphics[width=0.8\columnwidth]{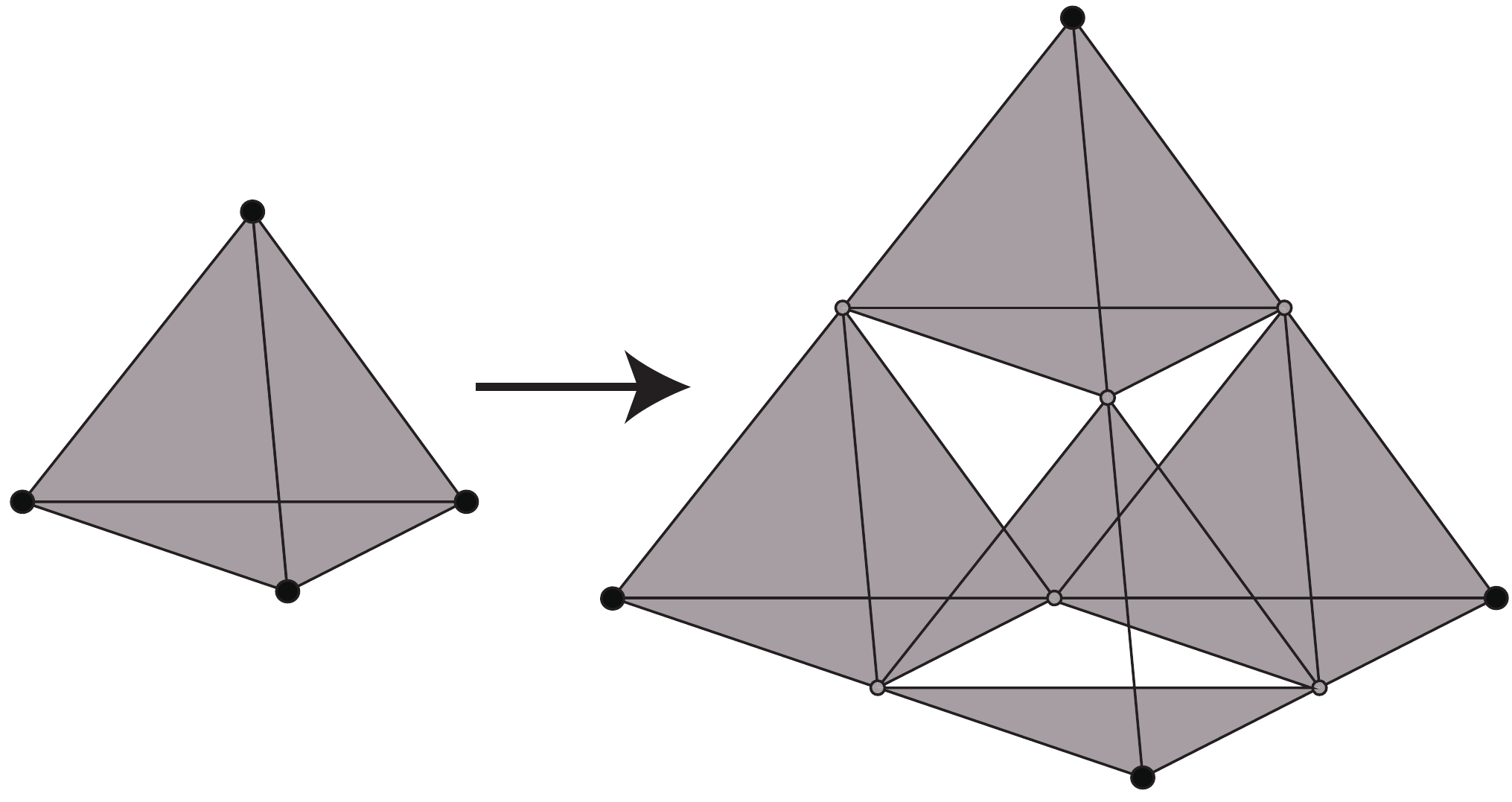}
	\caption{An illustration of fractal operators in Haah's code. The left figure shows a tetrahedron-like operator with 4 excitations (denoted by $\bullet$) in the corner. In the right figure, we pile up 4 tetrahedrons carefully, so that many excitations are cancelled (denoted by $\circ$)~\cite{Newman}, leaving a double-sized operator with 4 excitations. The process can be iterated. The operators we made has fractal structures and are called fractal operators.}\label{pic-fractal}	
\end{figure}

The analysis of Haah's code is a bit complicated. There are no string-like operators in Haah's code. The configuration with minimal number of excitations is be generated by a tetrahedron shaped fractal operator, resulting in 4 excitations in the corners. We can pile up 4 tetrahedrons as in Fig.~\ref{pic-fractal} so that many excitations are cancelled~\cite{Newman}. The result is another 4-excitation configuration with doubled length scale. The process can be iterated, until the length scale of the fractal operator reaches $L$, the length scale of the whole system. Thus, the number of 4-excitation configuration is roughly $N\ln N$. Here, $N$ is because we have a fractal operator with fixed size has $N$ different positions, $\ln N$ is because we have $\ln N$ different sizes. Thus, we have a term like $O(N\ln N)x^4$, which breaks down the low temperature expansion and indicates the zero-temperature phase transition. This again agrees with the analysis in Sec.~\ref{sec-example}.

Remarkably, the analysis for Haah's code is generalizable to general CSS codes. To simplify our discussion, we use the notions of topological charge and fractal generator~\cite{Haahmath}:
\begin{itemize}
	\item A \textbf{topological charge} is an excitation of finite energy which cannot be generated by finite Pauli operators but can be generated by infinite Pauli operators. An example is a single plaquette excitation in the 2D toric code. 
	\item A \textbf{fractal generator} is a way to copy and translate some topological charge so that the total configuration can be generated by finite Pauli operators. An example is the string operator in 2D toric code: after copy and move a plaquette excitation, we get a configuration with 2 plaquettes which can be generate by a string operator of finite length.
\end{itemize}
In Ref.~\onlinecite{Haahmath} it is proved that, for any three-dimensional, topological-ordered and degenerated (means degenerated ground states on some torus) Hamiltonian, there exists a fractal generator.

Given the existence of a fractal generator, we can do the same iteration as in Fig.~\ref{pic-fractal}, resulting in a series of fractal generators with different sizes and the same number of excitations. It contributes a $O(N\ln N)x^k$ term to the low temperature expansion to $Z(\beta)$. 

$k$ is not necessarily the minimal excitation though and one may suspect that the blowup coefficient $O(N\ln N)$ will be cured in $f\sim\ln Z(\beta)$. However, this will not happen. Recall that, as the Feynman diagram expansion, the expansion of $Z(\beta)$ corresponds to all graphs while the expansion of $\ln Z(\beta)$ corresponds to connected graphs. Fractal operators are always connected, otherwise part of the operator will generate the topological charge at some corner without generating other excitations, which contradicts the definition of topological charge. For example, a string operator in 2D toric code is connected, otherwise there must be other topological charges in the middle where the string breaks. 

In other word, we proved that the low temperature expansion of the free energy $f$ must break down, although not necessarily at the leading order.

\section{Conclusion and Discussion}\label{sec-conclusion}
In this article, we considered the topological entanglement entropy (TEE) for CSS codes at finite temperature. We find that the TEE is a constant within a non-critical phase. Therefore, assuming the criticality only appear at discrete points (i.e. no extended critical phase), the TEE is a piecewise constant function of temperature, with possible discontinuity only at critical points. Therefore, TEE serves as a good (nonlocal) order parameter for CSS codes.

One is therefore motivated to study the phase structure of CSS codes. For famous examples such as toric code, X-cube model, and Haah's code, we showed that they have a zero-temperature phase transition, due to the existence of a ``free sector", which allows us to calculate their partition functions explicitly. For general CSS codes, the ``free sector" does not exist in general, as we showed by a counterexample. However, based on the low temperature expansion, we can still argue the existence of a zero-temperature phase transition for CSS code with topological order in dimensions less or equal to 3.

Our results have promising connections to the problem of self-correction quantum memory. The self-correctness, or thermal stability, is the property that the mixing time under external perturbations (for example, thermal fluctuation in a bath) grows exponentially with system size~\cite{3dtoric,rev1}. It can be defined, for example, by considering the gap of the Lindblad operator $\mathcal{L}$ under a Markov environment~\cite{Alicki-2D,Alicki-4D}. Here, the inverse gap is roughly the mixing time. Although the self-correctness is by definition a dynamical property, it should have connections with the thermal Gibbs states $\rho_\beta=e^{-\beta H}/Z$, which is by definition a static object. Indeed, the Lindblad operator $\mathcal{L}$ can also be made Hermitian and $\rho_\beta$ is the ground state of  $\mathcal{L}$ \cite{Alicki-2D,Alicki-4D}. If $\mathcal{L}$ goes from gapped (i.e., the system is not self-correcting) to gapless (i.e., the system is self-correcting), then it should induce a transition in $\rho_\beta$, in the same way a gap-closing induces a quantum phase transition \cite{Sachdev}. 

For 2D toric code, X-cube model, and Haah's code, we have shown that the only phase transition is at $T=0$. Therefore, we believe that these models are not self-correcting for any nonzero temperature, as long as there is no ``purely dynamical" phase transitions that can be seen only from the spectrum of $\mathcal{L}$ but not the static partition function. For general 3D topological-ordered CSS codes, the phase transition at $T=0$ is a hint that they are not self-correcting, or at least a caveat that the beautiful topological memory scheme developed in zero temperature may not be used directly at finite temperature.

\begin{acknowledgements}
ZL would like to thanks Yuchi He for bringing the question into his attention and many fruitful discussions. This work is supported from NSF DMR-1848336. ZL is grateful to the Mellon fellowship and PQI fellowship at University of Pittsburgh. This research was supported in part by Perimeter Institute for Theoretical Physics. Research at Perimeter Institute is supported by the Government of Canada through the Department of Innovation, Science, and Economic Development, and by the Province of Ontario through the Ministry of Research and Innovation.
\end{acknowledgements}

\bibliography{finiteTO}

\appendix
\section{A code with no free sectors}\label{app-counter}
In Sec.~\ref{sec-example} we analyze the 2D and 3D toric code, X-cube model and Haah's code, and find that all models there contains a ``free sector", i.e., a sector that only have $o(N)$ global constraints when we use the stabilizers as elementary spins, thus behaves like decoupled spins, and thus a zero-temperature phase transition. In this appendix, we will see that in general this is not true by giving a counterexample.

Before going on, we need to comment on what we mean for a counterexample. 
\begin{enumerate}
\item A trivial ``counterexample" will be two decoupled 3D toric code model:
\begin{equation}\label{eq-stack}
	H=-\sum A_s(X)-\sum B_p(Z)-\sum A_s(Z')-\sum B_p(X'),
\end{equation}
i.e., we put two spins (with/without prime) on each link, use Pauli $X$ and Pauli $Z'$ to make star operators, and use Pauli $Z$ and Pauli $X'$ to make plaquette operators. For this model, both $X$ sector and $Z$ sector are not totally free: each contains a free subsector (star) and a nonfree subsector (plaquette), which is already enough for a zero-temperature phase transition. Therefore, what we mean by a counterexample is not ``neither $X$ sector nor $Z$ sector is equivalent to free spins", but actually ``does not contain a free \textbf{subsector}".
\item The star sector in the X-cube model is not free by definition, due to the local constraints $A_{xy}A_{yz}A_{zx}=1$, but can be reduced to independent terms by eliminating $A_{zx}$: $A_{zx}=A_{xy}A_{yz}$, which is still enough to ensure a zero-temperature phase transition. Therefore, by a counterexample we mean ``does not contain a subsector that can be reduced to independent spins by a \textbf{elimination} of some stabilizers".
\item The elimination can work for stabilizers in multiple sites. For example, one may consider a welding of 2D toric code and 2D Ising model:
\begin{multline}
    H=-\lambda \bigg(\sum_i A_i+\sum_i B_i\bigg)
    \\-\mu \bigg(\sum_{i,e}A_iA_{i+e}+\sum_{i,e}B_iB_{i+e}\bigg),
\end{multline}
where $e$ runs over two unit vectors $e_x,e_y$. The idea is, since $A_i$ (and $B_i$) for 2D toric code behaves like independent spins,  we can use them as spins in a 2D Ising model. One may hope that this will provide a code with no zero-temperature phase transition. However, this is not the case. Indeed, if $\lambda\neq 0$, this model is equivalent to a 2D Ising model in a nonzero magnetic field, which has only zero-temperature phase transition\footnote{The critical point in zero field $T_c$ is fragile because of the long range order.}. If $\lambda=0$, this model is thermodynamically good, but it's not a topological ordered system since the ground state can be distinguished by local measurements $A_i$.

From elimination point of view, new stabilizers $A_iA_{i+e}$ can be expressed from old stabilizers from two sites $A_i$ and $A_{i+e}$.

\end{enumerate}
\subsection{A short review of the commutative algebra formalism}
To explain our counterexample, we use the commutative algebra formalism. For more information, see ref.~\onlinecite{Haahmath}. For an introduction to commutative algebra, see ref.~\onlinecite{Atiyah}.

Denote $R$ be the Laurent polynomial ring $k[x,x^{-1},y,y^{-1},z,z^{-1}]$ where $k=\mathbb{F}_2$. Here, we have three variables $x,y,z$ because we work in 3D. Monomial $x^ay^bz^c$ corresponds to the point $(a,b,c)$ in $\mathbb{Z}^3$. A code can be represented by the following map:
\begin{equation}
\sigma: R^t\to R^{2q},
\end{equation}
where $R^n$ means free module over $R$ of rank $n$, $t$ is the number of stabilizer types, $q$ is the number of spins on each site. For example, for Haah's code, $n=q=2$. 

The rule for $\sigma$ is as follows. Each column corresponds to a stabilizer: if Pauli $X_i$ (or $Z_i$) operator for the $i^{\text{th}}$ spin at $(a,b,c)$ appears in that stabilizer, we add the monomial $x^ay^bz^c$ to the $i^{\text{th}}$ (or $(q+i)^{\text{th}}$ for $Z_i$) row of that column of $\sigma$. 
For example, for Haah's code,
\begin{equation}
\sigma=\begin{pmatrix} 
1+x+y+z& 0 \\
1+xy+yz+xz &   0  \\
0 & 1+\bar{x}+\bar{y}+\bar{z}  \\
0& 1+\bar{x}\bar{y}+\bar{y}\bar{z}+\bar{x}\bar{z}
\end{pmatrix},
\end{equation}
where the conjugate $\bar{x}=x^{-1}$. 

Define a symplectic structure on $R^{2q}$ by $\lambda_q=\begin{pmatrix} 
0 & 1_q \\
-1_q &   0 
\end{pmatrix}$, and denote $\epsilon=\sigma^{\dagger}\lambda_q$, then:

\noindent\textbf{Theorem} \cite{Haahmath} A Pauli stabilizer code has topological order iff $\ker\epsilon=\Ima\sigma$, or equivalently $\Ima\sigma=(\Ima\sigma)^{\perp}$ where the orthogonal completion is respect to the symplectic structure and the above conjugate.

\subsection{A nontrivial example with no free sectors}
For CSS codes, $\sigma$ has the form $\sigma=\begin{pmatrix} 
\sigma_1 &  \\
& \sigma_2 
\end{pmatrix}$, 
where $\sigma_i: R^{t_i} \rightarrow R^q$. The condition for topological order ($\Ima\sigma$ is a Lagrangian submodule in $R^{2q}$) is $(\Ima\sigma_1)^\perp=\Ima\sigma_2$ and $(\Ima\sigma_2)^\perp=\Ima\sigma_1$, where $\perp$ is now the Hermitian paring (with above conjugate). 

Whether the code has a free subsector after elimination is a property of $\Ima\sigma$ (indeed, the elimination process is equivalent to take $R$-linear combinations of the columns, with does not change $\Ima\sigma$). To be precise, the existence of free subsector is mathematically the existence of a free direct summand of $\Ima_i\sigma$ for $i=1\text{~or~}2$, i.e., a free submodule $M$ of $\Ima\sigma$ such that $\Ima\sigma_i=M\oplus M'$, where $M'$ is another submodule.

Denote $u=x+1,v=y+1,w=z+1$. Define $\sigma_i$ as follows ($t_1=3, t_2=9, q=8$, i.e., 8 qubits per site, 3 types of stabilizer made with Pauli $X$, 9 types of stabilizers made with Pauli $Z$):
\begin{equation}
\sigma_1=\begin{pmatrix} 
& uw &uv  \\
uw &    & u^2  \\
uv & u^2&  \\
& vw &v^2  \\
vw &    &uv  \\
v^2 & uv &0  \\
0 & w^2 &vw  \\
w^2 & 0 &uw
\end{pmatrix},
\overline{\sigma_2}=
\begin{pmatrix} 
v&u&&&w&v&&&  \\
&&&&&&&w&v\\
&w&v&&&&&&\\
&&&w&&u&&&\\
v&u&&&&&w&&u\\
w&&u&&&&&&\\
&&&v&u&&&&\\
&&&&&&v&u&\\
\end{pmatrix}.
\end{equation}
One can show that:
\begin{enumerate}
	\item $\Ima\sigma_1 \otimes K$ and $\Ima\sigma_2\otimes K$ are orthogonal complements of each other in $K^q$ where $K=\text{Frac}(R)$.
	\item $(\Ima\sigma_i\otimes K)\cap R^q$ is generated (as $R$-modules) by columns of $\sigma_i$. So we have $(\Ima\sigma_1)^\perp=\Ima\sigma_2$ and $(\Ima\sigma_2)^\perp=\Ima\sigma_1$.
\end{enumerate}
Therefore it is a code with topological order. 

One can check that $\ker\sigma_1=(u,v,w)^t$, $\ker\sigma_2=(\bar{u},\bar{v},\bar{w},0,0,0,0,0,0)^t\oplus(0,0,0,\bar{u},\bar{v},\bar{w},0,0,0)^t\oplus(0,0,0,0,0,0,\bar{u},\bar{v},\bar{w})^t$. This ensures that neither $\Ima\sigma_i$ contains a free direct summand. 

Indeed, consider $\Ima\sigma_1$ for example, we have
\begin{equation}
\Ima\sigma_1\simeq R^3/\ker\sigma_1.
\end{equation}
If $\Ima\sigma\simeq R\oplus M'$, then $\exists \mu: \Ima\sigma_1\to R$ which is surjective, then $\mu\sigma_1: R^3\to R$ is surjective, which must factorize through $\ker\sigma_1$.

Any surjective homomorphism $\mu\sigma_1: R^3\to R$ is given by a unimodular element $(f,g,h)$, i.e. $\exists p,r,s\in R$ such that $fp+gr+hs=1$. Moreover, the factorization implies $fu+gv+hw=0$. Let's multiply $f,g,h$ by their least common multiple of denominators and work in the polynomial ring $\mathbb{F}_2[x,y,z]=\mathbb{F}_2[u,v,w]$:
\begin{align}
&fp+gr+hs=x^ay^bz^c,\label{eq-unimodular}\\
&fu+gv+hw=0.
\end{align}
The second equation implies (consider the expansion as polynomials of $u,v,w$)
\begin{equation}
f=g'v+h'w.
\end{equation}
Plug into itself, we get $(g'u+g)v+(h'u+h)w=0$, which implies
\begin{align}
&g'u+g=f'w,\\
&h'u+h=f'v.
\end{align}
Evaluate it at $u=v=w=0$, we get $f=g=h=0$. Plug into \refeq{eq-unimodular}, we get a contradiction.

\end{document}